 \documentclass[sigconf]{acmart}
\acmSubmissionID{574}
\usepackage[ruled,vlined]{algorithm2e}
\RestyleAlgo{ruled}
\RestyleAlgo{vlined}

\acmJournal{TOG}

\usepackage{booktabs}

\newcommand{\bomega}{{\mathbf{\omega}}}
\newcommand{\bs}{\mathbf{s}}

\newcommand{\bd}{\mathbf{d}}

\SetKwComment{Comment}{// }{}

\def\figurePath{fig/}
\def\myfigure#1#2#3{
\begin{figure}[t]\centering\includegraphics[width = \linewidth]{\figurePath#2}\caption{#3}\label{fig:#1}
\end{figure}}

\def\mycfigure#1#2#3{\begin{figure*}[htb]\centering\includegraphics*[clip, width = \linewidth]{\figurePath#2}\caption{#3}\label{fig:#1}\end{figure*}}

\makeatletter
\newcommand\paragraphNew{\@startsection{paragraph}{4}{\parindent}%
  {-.5\baselineskip \@plus -2\p@ \@minus -.2\p@}%
  {-3.5\p@}%
  {\ACM@NRadjust{\@parfont}}}
	\makeatother
	
\AtBeginDocument{%
  \providecommand\BibTeX{{%
    \normalfont B\kern-0.5em{\scshape i\kern-0.25em b}\kern-0.8em\TeX}}}

\newcommand{\remind}[1]{\textcolor{black}{#1}}
\newcommand{\revise}[1]{\textcolor{black}{#1}}
\newcommand{\revises}[1]{\textcolor{black}{#1}}

\copyrightyear{2023} 
\acmYear{2023} 
\setcopyright{acmlicensed}\acmConference[SA Conference Papers '23]{SIGGRAPH Asia 2023 Conference Papers}{December 12--15, 2023}{Sydney, NSW, Australia}
\acmBooktitle{SIGGRAPH Asia 2023 Conference Papers (SA Conference Papers '23), December 12--15, 2023, Sydney, NSW, Australia}
\acmPrice{15.00}
\acmDOI{10.1145/3610548.3618198}
\acmISBN{979-8-4007-0315-7/23/12}

\begin{document}


\title{\revise{Multiple-bounce Smith Microfacet BRDFs using the Invariance Principle}}


\author{Yuang Cui}
\orcid{0009-0006-8983-7844}
\authornote{Contribute equally. Research done when Yuang Cui was an intern at Nanjing University of Science and
Technology. }
\affiliation{%
  \institution{Anhui Science and Technology University}
  \country{China}
}
\email{yuangcui@outlook.com}

\author{Gaole Pan}
\orcid{0009-0007-9335-333X}
\authornotemark[1]
\affiliation{%
  \institution{Nanjing University of Science and Technology}
  \country{China}
}
\email{pangaole@njust.edu.cn}

\author{Jian Yang}
\orcid{0000-0003-4800-832X}
\affiliation{%
  \institution{Nanjing University of Science and Technology}
  \country{China}
}
\email{csjyang@njust.edu.cn}

\author{Lei Zhang}
\orcid{0000-0002-2078-4215}
\affiliation{%
  \institution{The Hong Kong Polytechnic University}
  \country{China}
}
\email{cslzhang@comp.polyu.edu.hk}

\author{Ling-Qi Yan}
\orcid{0000-0002-9379-094X}
\affiliation{%
  \institution{University of California Santa Barbara}
  \country{USA}
}
\email{lingqi@cs.ucsb.edu}

\author{Beibei Wang}
\orcid{0000-0001-8943-8364}
\authornote{Corresponding author.}
\affiliation{
    \institution{Nankai University, Nanjing University of Science and Technology}
    \country{China}
}
\email{beibei.wang@njust.edu.cn}

\renewcommand{\shortauthors}{Cui et al.}

\begin{abstract}
\revise{Smith microfacet models are widely used in computer graphics to represent materials. Traditional microfacet models do not consider the multiple bounces on microgeometries, leading to visible energy missing, especially on rough surfaces. 
Later, as the equivalence between the microfacets and volume has been revealed, random walk solutions have been proposed to introduce multiple bounces, but at the cost of high variance. Recently, the position-free property has been introduced into the multiple-bounce model, resulting in much less noise, but also bias or a complex derivation.}
\revise{In this paper, we propose a simple way to derive the multiple-bounce Smith microfacet bidirectional reflectance distribution functions (BRDFs) using the invariance principle. At the core of our model is a shadowing-masking function for a path consisting of direction collections, rather than separated bounces.} \revise{Our model ensures unbiasedness and can produce less noise compared to the previous work with equal time, thanks to the simple formulation. Furthermore, we also propose a novel probability density function (PDF) for BRDF multiple importance sampling, which has a better match with the multiple-bounce BRDFs, producing less noise than previous naive approximations.}

\end{abstract}


\begin{teaserfigure}
\centering
\includegraphics[width=\textwidth]{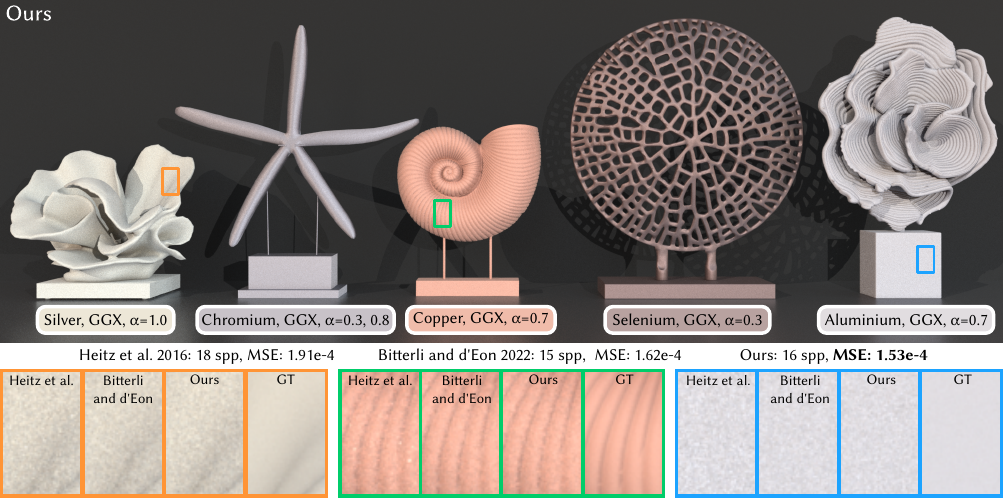}
\caption{We propose \revise{a multiple-bounce microfacet model derived with the invariance principle.} Our model produces results with less noise, compared to existing approaches (Heitz et al.~\shortcite{heitz2016} and Bitterli and d'Eon~\shortcite{BitterliAndd'Eon:2022}), with equal time (about 11.5 seconds for point/directional lighting and 100.0 seconds for environment lighting). Note that the bidirectional reflectance distribution function (BRDF) computation by Heitz et al.~\shortcite{heitz2016} is faster than others, so more samples are used for their method. \revise{Similarly, our method has a simpler formulation than Bitterli and d'Eon~\shortcite{BitterliAndd'Eon:2022}, resulting in more samples when rendered with equal time.}
}
\label{fig:teaser}
\end{teaserfigure}

\maketitle

\section{Introduction}
Material models are essential to realistic rendering, since they describe the interaction between light and surfaces. The microfacet model~\cite{Cook1982REFLEC, walter2007mmrt} is a commonly used analytical material model, which assumes that a surface is made of plenty of small microfacets. And the distribution of these microfacets, or normal distribution function (NDF), characterizes the principal appearance of the surface. Besides, the occlusion between microfacets is modeled using the shadowing-masking function, which encodes the proportion of light that may reach a microfacet without being occluded by others. Therefore, the classical microfacet bidirectional scattering distribution function (BSDF) can accurately model the \emph{single bounce} of light among the microfacets. However, since the light can actually bounce/scatter \emph{multiple times} before exiting the microgeometry, particularly for rough surfaces, the microfacet model can lead to an obvious energy loss. 

It is difficult to solve the multiple bounces of light from the microfacet model. Various assumptions have been made, and we are especially interested in the Smith approximation -- the NDFs are always the same regardless of different positions\revise{~\cite{smith:1967:smith}}. The Smith approximation immediately suggests the similarity between microfacets and a volumetric \revises{medium}, thus leading to random walk solutions by Heitz et al.~\shortcite{heitz2016} and Dupuy et al.~\shortcite{Dupuy:2016:Unification}. They trace light paths inside the volume, keeping track of the positions of each bounce and producing accurate results. \revise{Bitterli and d'Eon~\shortcite{BitterliAndd'Eon:2022} decrease the integral dimension using a closed-form formulation of the height distribution by assuming a hyperexponential distribution, leading to unbiased and less-noisy results. Wang et al.~\shortcite{wang2021positionfree} introduce the position-free property into the multiple-bounce computation, assuming the independence between the bounces (except sharing the same direction). Their model produces much less noise for both reflection and refraction and is unrelated to any specific height distribution. However, their results differ from the previous works due to the independent-bounce assumption.}

\revise{In this paper, we propose an unbiased multiple-bounce microfacet model with the invariance principle~\cite{amb1943},} which was introduced for planetary physics and radiative transfer by Ambartsumian~\shortcite{amb1943} for isotropic scattering and known as the Chandrasekhar's BRDF~\cite{Chandrasekhar1960}. We extend the invariance principle to handle anisotropic phase functions, resulting in a \emph{simple formulation} of the multiple-bounce Smith microfacet model. Thanks to this theory, we generalize the shadowing-masking function from a single bounce to an entire path built on top of the position-free multiple-bounce model by Wang et al.~\shortcite{wang2021positionfree}. \revise{In practice, since our model has a simpler formulation, it results in an even lower noise level compared to the previous method~\cite{BitterliAndd'Eon:2022} when rendered with equal time}, while \revises{retaining} their merits, e.g., passing the white furnace test, and working with anisotropic materials and general normal distributions such as Beckmann~\cite{BeckmannSpizzichino:1963} and GGX~\cite{walter2007mmrt}.

\revise{Furthermore, we improve the PDF of the multiple-bounce microfacet model by approximating the multiple bounces within a semi-infinite anisotropic \revises{medium} with an isotropic multiple-bounce approximation \cite{hapke1981}. Our PDF better matches the multiple-bounce function than the commonly used approximation~\cite{heitz2016, wang2021positionfree} --the single-bounce microfacet model together with a Lambertian term, leading to a higher quality when rendered with multiple importance sampling (MIS) \cite{VeachThesis}, particularly for low-roughness materials where the Lambertian approximation no longer holds.}

\revises{An open source implementation of our methods is available at https://github.com/wangningbei/sourceCodeMBBRDF.}

\begin{table}[!t]
	\renewcommand{\arraystretch}{1.1}
	\caption{\label{tab:notations} Notations.}
\begin{small}
  \begin{tabular}{|l|c|l|}\hline
  \multicolumn{2}{|c|}{Mathematical notation }\\
	\hline
		$\Omega$      			                                   &full spherical domain \\ 
		$\Omega^+$      			                                     &upper spherical domain \\ 
        $\Omega^-$      			                                     &lower spherical domain \\ 
        $\bomega_i \cdot \bomega_o$
                    &dot product \\
        $\left\vert \bomega_i \cdot \bomega_o \right \vert$
                    &absolute value of the dot product \\
        
         $o()$ & the infinitesimal of higher order \\
		\hline
		 \multicolumn{2}{|c|}{Physical quantities used in microfacet models}\\
				\hline
		$\omega_g = (0,0,1)$ & geometric normal \\
		$\omega_m $ & microfacet normal \\
		$\omega_i $ & incident direction \\
		$\omega_o $ & outgoing direction \\
		$\Lambda(\omega)  $ &the Smith Lambda function \\
		$D(\omega_m)$ &normal distribution function \\
		$D_{\omega_i}(\omega_m)$ &visible normals’ distribution \\
		$F(\omega_i,\omega_m)$ &Fresnel factor \\
		
		$\rho(\omega_i,\omega_o)$& multiple-bounce BSDF with the cosine term\\

		\hline
		\end{tabular}

\end{small}
\end{table}
\section{Related Work}
\label{sec:related}

In this section, we briefly review previous works related to multiple-bounce Smith microfacet model and the invariance principle.


\subsection{Multiple-bounce microfacet models.} 

The typical microfacet models only express the single bounce on the surface microgeometry, resulting in an energy loss. We group the existing works on multiple-bounce computations into two categories: physically-based and non-physically based models. 

\paragraph{Physically-based multiple-bounce microfacet models}
The multiple bounces of the Smith model does not have an explicit formula. Monte Carlo random walk has become one necessary solution for accurate results. For that, the microfacets are treated as randomly distributed microflakes~\cite{heitz2016} or homogeneous medium \cite{Dupuy:2016:Unification}, where Heitz et al.~\shortcite{heitz2016} need a height distribution function to trace the height\footnote{They observed that for \emph{a couple of specific} choices of height distributions (e.g., uniform distribution and Gaussian distribution), the rendered results are the same, without proof for generalization.}. These methods can be extended to normal-mapped surfaces~\cite{Schussler:2017:normal} or combined with wave optics~\cite{Falster:2020:Wave}.
Bitterli and d'Eon~\shortcite{BitterliAndd'Eon:2022} decrease the integral dimension by deriving an analytical preintegration of collision distances. The key insight of their model is an explicit formulation of the height distribution as a hyperexponential distribution. Their model produces identical results as Heitz et al.~\shortcite{heitz2016}, with less noise level. Recently, Wang et al.~\shortcite{wang2021positionfree} introduce the position-free property into the multiple-bounce computation, leading to significant variance reduction, by treating each bounce at the micro-scale separately. \revise{The independent bounce assumption in their model introduces bias, making their results different from Heitz et al.~\shortcite{heitz2016}. Similar to these works, our model follows the position-free assumption, reducing the multiple-bounce path integral dimensions to the angular dimension only. However, we remove the independent-bounce assumption. Thanks to the invariance principle, our model has a much simpler formulation and derivation.} And in practice, this leads to identical results to Bitterli and d'Eon~\shortcite{BitterliAndd'Eon:2022} but with even lower noise when rendered with equal time.

Different from Smith microfacet models, the V-groove models \cite{lEE2018PRATMUL, Feng2018multiV} allow for analytic solutions for multiple bounces. However, they produce shiny appearance even on rough surfaces and have singularities in the shadowing-masking term. We do not extend further discussion on these models.

\paragraph{Non-physically based multiple-bounce microfacet models}

Some existing methods~\cite{Xie:2019:multiple, Bai:2022:OT, Wang:2022:sponge} use neural networks to represent the multiple-bounce BSDFs, avoiding the random walk during rendering at the cost of introducing bias.

Some other approximate multiple-bounce models have been proposed for efficient rendering, by mixing the single scattering with an additional lobe empirically~\cite{KullaConty:2017:revisiting}, or by scaling the single bounce results~\cite{Turquin:2019:multiple}. These methods are fast, but are less accurate.

\subsection{The Invariance principle} 

Ambartsumian~\shortcite{amb1943} proposed the invariance principle for radiative transfer. He derived a semi-analytical formulation to model a multiple-bounce BRDF with an isotropic phase function within a semi-infinite medium. This is the first time that the invariance principle was used to model the light transport within a \revises{medium}. Later, Ambartsumian~\shortcite{amb1944} extended the semi-analytical formulation for arbitrary phase functions rather than isotropic ones. 

\sloppy
\revise{Chandrasekhar~\shortcite{Chandrasekhar1960} \revises{proposed} a different form of the multiple-bounce function, reducing to solving equations, known as $H$ functions. His model is also called \emph{Chandrasekhar's BRDF} in the following works.} 
Horak and Chandrasekhar~\shortcite{Horak1961} \revises{proposed} a three-term phase function with Legendre polynomials, leading to an analytical solution to the multiple bounces within a semi-infinite medium. However, the computation is highly complex, making rendering less practical. Recently, d'Eon~\shortcite{d2021analytic} introduced the three-term model by Horak and Chandrasekhar~\shortcite{Horak1961} for a Lambertian sphere phase function. By introducing some approximations, his model becomes much more practical and achieves a faithful appearance. \revise{Pharr and Hanrahan~\shortcite{PharrAndHanrahan:2000} use the invariance principle to derive the integral scattering equation for slabs of media. Note that their model still needs to sample the depth rather than angularly only, which is a key difference from ours.} 

Previous works cannot be applied to the multiple-bounce computation within the microfacets directly, since the phase function in this problem is anisotropic (depends on the direction). Therefore, we derive a novel scattering function specialized for our problem.

\section{Background and overview}

\myfigure{allModels}{localSmith.pdf}{These four models define the path space of multiple-bounce BRDFs in different ways. The path space defined by Heitz et al.~\shortcite{heitz2016} includes the height of the microgeometry. Bitterli and d'Eon~\shortcite{BitterliAndd'Eon:2022} turn the ray distance into the height difference and then perform preintegration on the spatial domain. Wang et al.~\shortcite{wang2021positionfree} and our model are position-free, except our model removes the independent-bounce assumption.}

In this section, we analyze the multiple-bounce Smith models related to ours, highlighting the differences between these methods. Then we provide an overview of our model.

\subsection{Path formulation of multiple-bounce Smith models}


Before analyzing the existing methods, we first define the notations. The light transport at any shading point $\bs$, potentially undergoing multiple bounces, is defined as a path integral for a given pair of query directions $\omega_i$ and $\omega_o$. Thus, the query directions $\omega_i$ and $\omega_o$ construct a path space. The light path is defined differently in the existing approaches, as shown in Figure~\ref{fig:allModels}. Nevertheless, this light path includes two domains: a spatial domain and an angular domain. We use a set of directions to define the angular domain: $\bar x_a = \left ({\bd_0,\bd_1,\dots,\bd_k} \right)$. The first and last directions are aligned with the macro incident and outgoing directions of a BSDF query, i.e., $\bd_0=\omega_i$ and $\bd_k=\omega_o$. Note that the direction of $\omega_i$ points downwards. About the spatial domain, different models have different parameterizations.

Heitz et al.~\shortcite{heitz2016} treat the microfacets as a microflake volumetric model, leading to a path defined as $\bar x_\mathrm{H} =(h_0, \bd_0, h_1, \bd_1,\dots,\bd_k)$. Dupuy et al.~\shortcite{Dupuy:2016:Unification} simplify the medium from anisotropic media to a homogeneous media, resulting in a path denoted as $\bar x_\mathrm{D} = (\bd_0, t_0, \bd_1, t_1,\dots,\bd_k)$. Recently, Bitterli and d'Eon first transform the distance along a ray to a height difference $\Delta z$ and then perform a preintegration of the height component; thus, the final path of their model $\bar x_\mathrm{B} = (\bd_0, \bd_1,\dots,\bd_k)$, remains in the angular domain only. The variable change from the ray distance to the height difference allows for integrating explicitly over the depths, resulting in lower variance. All the above models provide the same result. Another model by Wang et al.~\shortcite{wang2021positionfree} introduces the position-free property into the micro scale of a BRDF, resulting in the angular only domain directly $\bar x_\mathrm{W} = (\bd_0, \bd_1,\dots,\bd_k)$. 


\subsection{Position-free multiple-bounce Smith microfacet model}
Since our model is under the position-free framework by Wang et al.~\shortcite{wang2021positionfree}, we briefly review their model. Wang et al.~\shortcite{wang2021positionfree} introduced the position-free property into the micro scale of a BRDF, resulting in a simple multiple-bounce solution for Smith microfacet models.

Starting from the light path $\bar x = \left ({\bd_0,\bd_1,\dots,\bd_k} \right)$, the contribution $f(\bar x)$ of this light path is the product of vertex terms $v_i$ (on each vertex) and segment terms $s_i$ (on each direction):
\begin{equation}
    f(\bar x) = \prod_{i=0}^{i=k-1} v_i s_i.
    \label{eq:Wang}
\end{equation}

The vertex term $v_i$ is defined to represent local interactions between the light and the microfacets, consisting of the normal distribution function $D$, the Fresnel term $F$, and the Jacobian term:
\begin{equation}
v_i =  \frac{F \left(-\bd_i, \omega_h^{i}\right) D \left(\bomega_h^{i} \right)}{4 \left\vert \bomega_g \cdot \left(-\bd_i \right) \right \vert},
\label{eq:v_r}
\end{equation}
where $\omega_g$ is the macrosurface normal and $\omega_h^{i}$ denotes the half vector between $\bd_i$ and $\bd_{i+1}$.

The segment term considers the shadowing-masking function of the light path. Two types of shadowing-masking functions are proposed by Wang et al.~\shortcite{wang2021positionfree}: a height-uncorrelated shadowing-masking and a height-correlated one. Both functions are defined on each bounce separately. The latter considers the correlation between incoming and outgoing rays at each bounce, while the former treats the incoming and outgoing rays at each bounce independently. The final segment terms are the accumulation of each bounce. This bounce-independent assumption leads to different rendered results from the other models.



\subsection{Overview}

We follow the position-free property by Wang et al.~\shortcite{wang2021positionfree}, but with one essential difference: considering the correlation of the segment term among bounces. We notice that the segment term of a bounce depends on the previous bounces. Therefore, the segment term should be defined for the entire light path rather than for individual bounces. \revise{In this way, the contribution of a light path with bounce $k$ in our model consists of the vertex terms and a segment term $S_k$ defined on the light path:}
\revise{\begin{equation}
    f(\bar x) = \left(\prod_{i=0}^{i=k-1} v_i\right) S_k(\bd_0,\bd_1,\dots,\bd_k),
    \label{eq:lightpath}
\end{equation}}
where $v_i$ is identical as Eqn.~(\ref{eq:v_r}). 



In the next section, we \revise{derive a segment term $S_k$ for a light path} in an elegant way using the invariance principle. Once we have the formation of $S_k$ and thus $f(\bar x)$ for a single light path, the final multiple-bounce BRDF can be computed similar to Wang et al.~\shortcite{wang2021positionfree}. We just need unidirectional path tracing or bidirectional path tracing, both position-free, to sample in the path space $\bar x_\mathrm{W} = (\bd_0, \bd_1,\dots,\bd_k)$, where each sampled light path consists of only sampled directions.

\section{The path segment term via the invariance principle}
\label{sec:method}

\sloppy
In this section, we derive our \revise{segment term} for a sampled light path using the invariance principle~\cite{amb1943} (Sec.~\ref{sec:reflection}). Then we analyze the main properties and merits of our model (Sec.~\ref{sec:properties}). 

The invariance principle is based on the evident but insightful fact that adding a layer of small optical thickness $\Delta\tau$ with the same properties as the original medium to this medium, should not change its reflectivity. It implies that the total contribution of the processes associated with the added layer must be zero. We follow this principle, and introduce it for \revises{medium} with infinite thickness, resulting in our multiple-bounce formulation.

\subsection{Multiple-bounce formulation for the reflective microfacet model}
\label{sec:reflection}

\myfigure{cases}{fig3.pdf}{Four cases of light transport due to the adding of the thin layer: (a) \revises{extinction} only within the thin layer; (b) scattering once within the thin layer without reaching the bottom medium ; (c) scattering at the incoming ray, and (d) scattering at the outgoing ray.}

We follow the assumption of Dupuy et al. ~\shortcite{Dupuy:2016:Unification} and Bitterli and d'Eon~\shortcite{BitterliAndd'Eon:2022} that the microfacet model is equivalent to a semi-infinite homogeneous \revises{medium}. Given a semi-infinite homogeneous \revises{medium} $\mathcal{M}$, we add a slice of \revises{medium} with thickness $\Delta\tau$ on top of $\mathcal{M}$. The addition of this thin layer causes some changes in the light transport within the \revises{medium}. As shown in Figure~\ref{fig:cases}, we recognize four cases:
\begin{enumerate}
    \item \revises{extinction} within this thin layer,
    \item scattering at thin layer without reaching the \revises{medium} $\mathcal{M}$,
    \item scattering within the incoming ray when crossing the thin layer and then scattering within the \revises{medium}, and leave from the outgoing ray from the thin layer without scattering, and
    \item the symmetric case of case 3.
\end{enumerate}

We only consider no scattering (case 1) or scattering once (the other three cases) in the thin layer. Since $\Delta\tau$ is chosen to be sufficiently thin, which can be at least an order of magnitude thinner than the mean free path, there is no need to consider higher-order scattering events.


\paragraph{\textsc{Case 1.}} The new added layer makes attenuation on both the incoming and the outgoing directions, leading to the following change in the outgoing radiance:
\begin{equation}
    \begin{aligned}
\Delta L_{0} & =\rho\left(\omega_{i}, \omega_{o}\right) L\left(1-e^{-\Delta \tau \left(\left|\Lambda\left(\omega_{i}\right)\right|+\Lambda\left(\omega_{o}\right)\right)}\right) \\
& =\rho\left(\omega_{i}, \omega_{o}\right)\revises{\left(L \Delta \tau \right)}\left(\left|\Lambda\left(\omega_{i}\right)\right|+\Lambda\left(\omega_{o}\right)\right) +o(\Delta \tau ),
\end{aligned}
\end{equation}
where $\rho(\omega_i,\omega_o)$ represents the multiple-bounce BRDF with the cosine term. $L(\omega_i)$ is the incoming radiance, written as $L$ for clarity. When $x$ is small enough (but still a positive number), $1-e^{-x}$ will converge to $x$. \revises{The infinitesimal of higher order is denoted by $o(\Delta \tau )$}, which can be ignored. $\Lambda$ is the Smith Lambda function~\revises{\cite{Heitz2014Microfacet}}. 


\paragraph{\textsc{Case 2.}} The light is scattered only in the added layer, without reaching the \revises{medium} $\mathcal{M}$:
\begin{equation}
    \begin{aligned}
\Delta L_{1} & =f_p\left(\omega_{i}, \omega_{o}\right) L\left(1-e^{-\Delta \tau\left|\Lambda\left(\omega_{i}\right)\right|}\right)e^{-\Delta \tau \Lambda\left(\omega_{o}\right)} \\
& =f_p\left(\omega_{i}, \omega_{o}\right)\revises{\left(L \Delta \tau \right)}\left|\Lambda\left(\omega_{i}\right)\right|+o(\Delta \tau )\\
& =v\left(\omega_{i}, \omega_{o}\right)L \Delta \tau  +o(\Delta \tau ),
\end{aligned}
\end{equation}
where $\Delta L_{1}$ is the change of the outgoing radiance due to the added layer. Note that $e^{-x}$ converges to 1 for a small positive number $x$. \revises{The phase function $f_p$ of the scattering}, as Heitz et al.~\shortcite{heitz2016}, defined as:
\begin{equation}
    f_p =  \frac{F \left(\omega_i, \omega_h\right) D_{\omega_i} \left(\omega_h \right)}{4 \left\vert \omega_h \cdot \omega_i \right \vert},
\end{equation}
where $D_{\omega_i}$ is the visible normal distribution function (VNDF). Interestingly, the product of the phase function and the $\left|\Lambda\left(\omega_{i}\right)\right|$ results in the vertex term of the single bounce.

\paragraph{\textsc{Case 3.}} The light is scattered once in the added layer, reaches \revises{medium} $\mathcal{M}$, and later leaves the \revises{medium} by going through the thin layer. Since the scattering direction in the added layer could be any direction in the lower-hemisphere, there is an integral on the lower-hemisphere here: 
\begin{equation}
    \begin{aligned}
\Delta L_{2} 
& =\int _{\Omega^-}f_p\left(\omega_{i}, \omega\right)\revises{\left(L \Delta \tau \right)}\left|\Lambda\left(\omega_{i}\right)\right|\rho \left(\omega, \omega_{o}\right)\mathrm{d}\omega+o(\Delta \tau )\\
& =L \Delta \tau\int _{\Omega^-}v\left(\omega_{i}, \omega\right)\rho \left(\omega, \omega_{o}\right)\mathrm{d}\omega+o(\Delta \tau ).
\end{aligned}
\end{equation}
\paragraph{\textsc{Case 4.}} The symmetric case of case 3, where the scattering happens on the outgoing direction:
\begin{equation}
    \begin{aligned}
\Delta L_{3} 
& =\int _{\Omega^+}\rho \left(\omega_{i}, \omega\right)\revises{\left(L \Delta \tau \right)}\left|\Lambda\left(\omega\right)\right|f_p\left(\omega, \omega_{o}\right)\mathrm{d}\omega+o(\Delta \tau )\\
& =L \Delta \tau\int _{\Omega^+}\rho\left(\omega_{i}, \omega\right) v\left(\omega, \omega_{o}\right)\mathrm{d}\omega+o(\Delta \tau ).
\end{aligned}
\end{equation}
The invariance principle implies that the four cases should guarantee energy conservation:
\begin{equation}
    \Delta L_{0} =\Delta L_{1}+\Delta L_{2}+\Delta L_{3}.
\end{equation}
Then, we have
\begin{equation}
    \begin{aligned}
&\rho\left(\omega_{i}, \omega_{o}\right) \revises{\left(L \Delta \tau \right)}\left(\left|\Lambda\left(\omega_{i}\right)\right|+\Lambda\left(\omega_{o}\right)\right)=v\left(\omega_{i}, \omega_{o}\right)L \Delta \tau\ +\\
&L \Delta \tau\int _{\Omega^-}v\left(\omega_{i}, \omega\right)\rho \left(\omega, \omega_{o}\right)\mathrm{d}\omega+L \Delta \tau\int _{\Omega^+}\rho\left(\omega_{i}, \omega\right) v\left(\omega, \omega_{o}\right)\mathrm{d}\omega.
\end{aligned}
\end{equation}
\revises{By cancelling some common factors, and unifying the integral domain to the upper hemisphere, we have}
\begin{equation}
    \begin{aligned}
&\rho\left(\omega_{i}, \omega_{o}\right)= \frac{1}{\Lambda\left(-\omega_{i}\right)+\Lambda\left(\omega_{o}\right)+1}\bigg(v\left(\omega_{i}, \omega_{o}\right)+\\
&\int _{\Omega^+}\Big( v\left(\omega_{i}, -\omega\right)\rho \left(-\omega, \omega_{o}\right)+\rho\left(\omega_{i}, \omega\right) v\left(\omega, \omega_{o}\right)\Big) \mathrm{d}\omega \bigg).
\end{aligned}
\end{equation}

Starting from the above equation, we can easily derive the contribution of a sampled light path $\bar x^{0}_k = (\bd_0,\bd_1,\dots,\bd_k)$ with bounce $k$:


\begin{equation}
f\left( \bar x^{0}_{k} \right)=  \frac{v\left(\bd_{0}, -\bd_{k-1}\right) f \left(x^{1}_{k}\right)+  v\left(\bd_{k-1}, \bd_{k}\right) f\left(x^{0}_{k-1}\right)}{\Lambda\left(-\bd_{0}\right)+\Lambda\left(\bd_{k}\right)+1} .
\end{equation}

Substituting Eqn.~(\ref{eq:lightpath}) into the above equation and canceling the vertex terms leads to the segment term of a light path :
\begin{equation}
\begin{aligned}
S_k\left(\mathbf{d}_0, \dots , \mathbf{d}_k\right) = S_1\left(\mathbf{d}_0,  \mathbf{d}_k\right)\left ( S{_{k-1}}\left(\mathbf{d}_0, \dots ,  \mathbf{d}_{k-1}\right)+\right.\\
\left.S_{k-1}\left(\mathbf{d}_1, \dots ,  \mathbf{d}_k\right) \right ).
\label{eq:segmentPathCase}
\end{aligned}
\end{equation}

\myfigure{threebounce}{pathtype_v2.pdf}{Three types of configurations in terms of the ray directions for a light path with three bounces. The first path type has the second recursive term in Eqn.~(\ref{eq:segmentPathCase}) only, and the second one has the first recursive term in Eqn.~(\ref{eq:segmentPathCase}) only, and the last path type has both recursive terms. The outgoing ray pointing downwards, or the incoming ray coming from the lower hemisphere indicates that the light path is invalid.}

The above equation shows that the segment term of a light path with bounce $k$ consists of two recursive terms. However, we find that only a single recursive term has a physical meaning for some light paths. For example, in Figure~\ref{fig:threebounce}, we show three path configurations (the orientation of the ray) for a path with three bounces. Only one recursive term exists for the first and second path types. The other recursive term in these two types has an invalid incoming/outgoing ray direction, i.e., the outgoing ray pointing downwards or the incoming ray coming from the lower hemisphere. For the last path configuration, both recursive terms exist. We also need to formulate these rules into our formulation: 

\begin{equation}
    \begin{aligned}
S_1\left(\mathbf{d}_0,  \mathbf{d}_1\right) & =\frac{1}{1+\Lambda (-\mathbf{d}_0)+\Lambda (\mathbf{d}_1)},  \\
S_k\left(\mathbf{d}_0, \dots , \mathbf{d}_k\right) & = \begin{cases}
0,\text { if } \mathbf{d}_{0} . z>0\text { or }\mathbf{d}_{k} . z<0,\\
S_1\left(\mathbf{d}_0,  \mathbf{d}_k\right)\left ( S{_{k-1}}\left(\mathbf{d}_0, \dots ,  \mathbf{d}_{k-1}\right)+\right.\\
\left.S_{k-1}\left(\mathbf{d}_1, \dots ,  \mathbf{d}_k\right) \right ),\text{ otherwise}. \\
\end{cases}\\
\end{aligned}
\label{eq:final}
\end{equation}

In the above equation, we find that our single-bounce segment term ($S_1\left(\mathbf{d}_0, \mathbf{d}_1\right)$) also matches the height-correlated shadowing-masking function by Ross et al.~\shortcite{Ross:05:Lambda}. \revise{The implementation of our model is detailed in Sec.~\ref{sec:impl}.} 

\revise{In theory, the time complexity of our method is $O(NM)$, where $N$ and $M$ represent the ray count pointing upwards and downwards respectively and $N+M==k$. Therefore, the best case of the time complexity is linear \revises{in} $k$, and the worst case is quadratic \revises{in} $k$. However, in practice, up to 10 bounces, the time cost is very close to a linear function of $k$, since most of the light paths (about 90\% for Vase scene\revises{, as shown in Figure \ref{fig:linegraph}}) only have one or two directions pointing downwards. In Figure~\ref{fig:linegraph}, we show \remind{the segment term computation time cost} curve as a function of the bounce count and provide an in-depth discussion.} 

\revise{Bitterli and d'Eon~\shortcite{BitterliAndd'Eon:2022} proposed a similar concept $p_\mathrm{exit}$, which is the probability of a photon exiting the medium, conditioned on the directions it takes after each collision. This $p_\mathrm{exit}$ is similar but not equivalent to our segment term, since some terms in their $p_\mathrm{exit}$ are included in our vertex term. \remind{We provide an in-depth discussion in the supplementary.}}

\subsection{Properties and analysis}
\label{sec:properties}

We analysis the main properties of our model in this section. 

\paragraph{Unbiasedness and reciprocity.} 
Our model is unbiased and produces the identical results as Heitz et al.~\shortcite{heitz2016}. We provide the convergence curve of our model w.r.t. varying sample rate in Figure~\ref{fig:linegraph}. \revise{The removal of the independent-bounce assumption from Wang et al.~\shortcite{wang2021positionfree} makes our model unbiased.} Furthermore, our model is reciprocal, since both our vertex terms and the segment term Eqn.~(\ref{eq:final}) are reciprocal.

\paragraph{\revise{Relationship to the model by Bitterli and d'Eon \shortcite{BitterliAndd'Eon:2022}}}
\revise{
Both Bitterli and d'Eon~\shortcite{BitterliAndd'Eon:2022} and our model are unbiased. We also find that their model and ours are equivalent for a specific bounce (e.g., bounces = 2 or 3) with a non-trivial derivation, \revises{as shown in (Sec. 1.2) (supplementary)}. However, a general derivation from their model to ours for an arbitrary bounce is not apparent. Our model shows two benefits compared to theirs. First, the derivation of Bitterli and d'Eon~\shortcite{BitterliAndd'Eon:2022} depends on the height distribution and results in a more complex formulation, while our model has a simpler derivation thanks to the invariance principle, leading to time efficiency. Second, our model avoids singularities, which come from the minus of two $\Lambda$ functions \revises{as shown in Eqn.~(6) (supplementary), although the numerical stability is rarely an issue for BRDF evaluation.} Hence, the rendered results of Bitterli and d'Eon~\shortcite{BitterliAndd'Eon:2022} have the same variance as ours with an equal number of samples, but are noisier rendered with equal time. In the supplementary, \remind{we also provide the algorithms for both models, and explicitly show the reason for our time efficiency}. }

\subsection{Efficient BRDF evaluation and sampling}
\label{sec:impl}

\revise{In this section, we show the implementation details of the two key components for a BRDF: evaluation and sample.}

\revise{\paragraph{Evaluation.}
We provide two estimators for BRDF evaluation: a unidirectional estimator using path tracing and a bidirectional estimator using bidirectional path tracing. We provide the details for the unidirectional estimator only, and the bidirectional version is the same as Wang et al.~\shortcite{wang2021positionfree}, except for the segment term.}

\revise{Starting from the incoming ray direction $\bd_0$, we sample the visible normal distribution function to get a new direction and perform the next event estimation by connecting with the outgoing direction $\bd_k$. \revises{The path sampling termination is controlled by the Russian roulette~\cite{Arvo:1990}}. To compute $S_k$ efficiently, we use dynamic programming, as shown in Alg. 1 ( supplementary). }


\SetKwProg{Class}{class}{}{end}
\SetKwProg{Fn}{function}{}{end}
\SetKw{Def}{def}
\DontPrintSemicolon

\newcommand{\mycommfont}[1]{\small\texttt{\textcolor[HTML]{4D8F33}{#1}}}
\SetCommentSty{mycommfont}

\revise{\paragraph{Sample.}
At each bounce, we sample the VNDF to get the outgoing direction and \revise{then compute the masking function ($G_1$ function) of the sampled ray to decide whether to exit the microgeometry}. Otherwise, we treat the sampled direction as the incoming direction and continue sampling until the ray leaves the surface. After getting such a light path, we compute the path contribution by evaluating the light path divided by the PDF \revises{of the sampled path.}}

\revise{\subsection{Improved PDF for multiple-bounce BRDF} 
 The PDF of a multiple-bounce BRDF is important for multiple importance sampling. Since no analytical formulation exists for multiple-bounce BRDFs, the previous works~\cite{heitz2016,wang2021positionfree} estimate the multiple-bounce PDF with a Lambertian term and a single-bounce function, which can fit the actual multiple-bounce BRDFs well for high-roughness materials, but tends to overestimate the PDF at grazing angle directions for low-roughness BRDFs.}
 
 \revise{Our key insight is that although an anisotropic \revises{medium} has no analytical multiple-bounce formulation, it can be estimated by an isotropic multiple-bounce formulation derived by Hapke~\shortcite{hapke1981}:}
 \revise{\begin{equation}
\begin{aligned}
f_{\mathrm{mul}}(\omega_i,\omega_o) & = \frac{a}{4\pi}\frac{H\left(\left\vert \omega_g \cdot \omega_i \right \vert\right)H\left(\left\vert \omega_g \cdot \omega_o \right \vert\right)-1}{\left\vert \omega_g \cdot \omega_i \right \vert+\left\vert \omega_g \cdot \omega_o \right \vert},
 \\ H(\mu) & = \frac{1+2\mu}{1+2\sqrt{1-a}\mu},
\end{aligned}
\end{equation}
where $a$ is the \revises{albedo of the medium}.}

\revise{We construct such an isotropic \revises{medium} from the \revises{microfacet}, by setting the \revises{albedo of the medium} as $a = \frac{\alpha_x+\alpha_y}{2}$. Note that this mapping is empirical, due to the observation that the \revises{albedo of the medium} dominates the scattering distribution for an isotropic \revises{medium} and the roughness dominates the scattering distribution for the microfacet model.
Our final PDF includes the single bounce term and the estimated multiple bounce term:
\begin{equation}
\begin{aligned}
f_{\mathrm{PDF}}(\omega_i,\omega_o) & = \frac{D_{\omega_i} \left(\omega_h \right)}{4 \left\vert \omega_h \cdot \omega_i \right \vert}+f_{\mathrm{mul}}(\omega_i,\omega_o).
\end{aligned}
\end{equation}}

\revise{Our PDF matches the ground truth better than the previous estimation combining the single bounce and the Lambertian term, as shown in Figure~\ref{fig:pdfCurve} for both low-roughness and high-roughness materials. Note that our PDF is less accurate for anisotropic BRDFs than for isotropic \revises{medium}, since our mapping function averages the roughness for anisotropic BRDFs. \revises{Our PDF still shows better matching with the ground truth than the single bounce + the Lambertian term approximation, \revises{as shown in Figure 1 (supplementary). Note that an approximated PDF for multiple importance sampling does not introduce bias.}} }

\section{Results}
\label{sec:results}


We have implemented our algorithm inside the Mitsuba renderer \shortcite{Mitsuba}. The implementations of Heitz et al.~\shortcite{heitz2016} and Wang et al.~\shortcite{wang2021positionfree} are from the \revises{authors' websites}. \revises{We use the height-correlated version of Wang et al.~\shortcite{wang2021positionfree}.} We also reimplemented Bitterli \revises{and d'Eon}~\shortcite{BitterliAndd'Eon:2022} (PT). \revise{We did not provide the results of Bitterli \revises{and d'Eon}~\shortcite{BitterliAndd'Eon:2022} (BDPT), since we \revises{were not able to} 
reimplement their model with their provided algorithm. A detailed discussion is shown in the supplementary.} For the \revises{ground truth images in Figures~\ref{fig:teaser},~\ref{fig:sphere} and ~\ref{fig:vase}}, we refer to the converged results using Heitz et al.~\shortcite{heitz2016}. We use MSE to measure the difference between each method and the ground truth. All timings in this section are measured on a 2.20GHz Intel i7 (48 cores) with 32 GB of main memory. 

\subsection{Comparisons}

In Figure~\ref{fig:curve}, we compare the mean BRDF reflectance and the inverse efficiency as a function of the outgoing direction for a given incoming direction over several approaches, including our method, \revise{Wang et al.~\shortcite{wang2021positionfree}}, Heitz et al.~\shortcite{heitz2016}, and Bitterli and d'Eon~\shortcite{BitterliAndd'Eon:2022}. The inverse efficiency means the product of the variance and time cost, and a lower value indicates a higher quality. We show two roughnesses ($\alpha = 0.5$ and $1.0$) and two incoming directions. Our BRDF value can match both Heitz et al.~\shortcite{heitz2016} and Biltterli and d'Eon~\shortcite{BitterliAndd'Eon:2022}, while  our method is the most efficient in all cases (varying roughness or different incoming directions). \revise{Wang et al.~\shortcite{wang2021positionfree} \revises{cannot} match the ground truth and introduce bias due to the independent-bounce assumption, although their model has higher performance than ours, when $\theta_o$ = 0 for $\alpha = 1.0$.}
 
\paragraph{Matpreview scene.} 
In Figure~\ref{fig:sphere}, we show a copper Matpreview scene (GGX model, $\alpha$ = 1.0) lit by a directional light. We compare our models (PT and BDPT), Wang et al.~\shortcite{wang2021positionfree} (PT and BDPT) and Bitterli and d'Eon~\shortcite{BitterliAndd'Eon:2022} (PT) with equal sampling rate (4 sample per pixel
(spp)). Our model (BDPT) outperforms the other methods with a slightly longer time. The result of Bitterli and d'Eon~\shortcite{BitterliAndd'Eon:2022} has the same variance as ours (PT) with an equal sampling rate, but has a higher time cost due to the complex formulation. Wang et al.~\shortcite{wang2021positionfree} produce results with a low noise level, but with higher variance, since their model is biased. \revise{We also provide the difference images, which clearly show that ours (BDPT) has the lowest error.}

In Figure~\ref{fig:linegraph}, we show the MSE curve in terms of sampling rate for our methods (BDPT and PT), Heitz et al.~\shortcite{heitz2016} and Wang et al.~\shortcite{wang2021positionfree} on the Matpreview scene. The ground truth is the converged result of Heitz et al.~\shortcite{heitz2016}. Our methods (BDPT and PT) can converge to the reference, which indicates that our model is unbiased. On the contrary, the model by Wang et al.~\shortcite{wang2021positionfree} is biased due to the independent-bounce assumption. Our BDPT version has the fastest convergence compared to the others.   

\paragraph{Vase scene.} 
Figure~\ref{fig:vase} shows three statues (copper (GGX, \revise{$\alpha$ = 0.1}), aluminum (Beckmann, $\alpha$ = 0.6), and gold (GGX, $\alpha$ =0.5)) on a diffuse floor with direct lighting only, lit by an environment map and a point light. We use 256 spp for the environment map lighting to better show the effect of the BRDF evaluation. To achieve equal time, we use 17 spp for our method, 34 spp for Heitz et al.~\shortcite{heitz2016} and 23 spp for Wang et al.~\shortcite{wang2021positionfree} for the point light source. The rendered result by Heitz et al.~\shortcite{heitz2016} has the highest noise level. Although Wang et al.~\shortcite{wang2021positionfree} has a similar variance as ours, their result has a larger error because of the fundamental bias.  

\paragraph{DecorativeSet scene. }
We also provide a DecorativeSet scene with several statues in Figure~\ref{fig:teaser}, including different roughness as shown in the figure. In this scene, the statues are lit by an environment map, a directional light, and a point light, considering both direct lighting and indirect lighting. We compare our method against other two methods (Heitz et al.~\shortcite{heitz2016} and Bitterli and d'Eon~\shortcite{BitterliAndd'Eon:2022}) with equal time and use the unidirectional estimator for all the methods. Since the BRDF sample is not mentioned in Bitterli and d'Eon~\shortcite{BitterliAndd'Eon:2022} and the source code is not available, we use the same BRDF sample as Heitz et al.~\shortcite{heitz2016}. We use 128 spp for the environment map lighting to better show the effect of the BRDF evaluation. To achieve equal time, the sampling rates are set differently for different methods. Our method outperforms the other methods visually and quantitatively for all the settings.


\revise{\paragraph{Pot scene. }
We validate the impact of our improved PDF by showing PDF curves in Figure~\ref{fig:pdfCurve} and rendered results in Figure~\ref{fig:pdf_pot}, comparing with the approximation (single bounce + the Lambertian term) by previous work \cite{heitz2016, wang2021positionfree}. Figure~\ref{fig:pdf_pot} shows three aluminium pots of different roughnesses ($\alpha$ = 0.15, $\alpha$ = 0.2 and $\alpha$ = 0.1) with direct lighting from environment map. We use our BRDF evaluation and sample for both results and only differ in the PDF function. By comparison, our PDF produces less noise than the previous approximation.}


\revise{\subsection{Performance analysis}
In Figure~\ref{fig:linegraph}, we show the time cost of the \emph{segment term computation} \revises{w.r.t.} the bounce count. We find that the time cost is almost linear as the bounce count, although our model has quadratic time complexity in theory. The time cost depends on the ray counts, which point upwards and downwards. We find that only a tiny fraction of the cases have almost even downward and upwards ray counts \revises(as shown in the right curve) when the time complexity is high. This explains that the actual time cost is almost linear, although the theory time complexity is quadratic.}

\subsection{Discussion and limitations}

We introduce the invariance principle in the multiple-bounce microfact model. Unfortunately, our formulation is only derived for the spatial domain. Like any other existing work in this line of research, a Monte carlo estimator is still required for the angular domain to get a path.

Introducing the invariance principle to the multiple bounces of a refractive surface is much more complex than the reflective case. Our model relies on the equivalence between the microfacet and a homogeneous \revises{medium}. However, a well-known problem is that for a refractive surface, this equivalence leads to a discontinuous mapping in the spatial domain. For the same reason, Bitterli and d'Eon~\shortcite{BitterliAndd'Eon:2022} also consider reflective surfaces only.

\section{Conclusion}

In this paper, we present a multiple-bounce BRDF model. \revise{Theoretically, we introduce a novel way -- the invariance principle, to derive a multiple-bounce BRDF;} practically, we propose an analytical formulation for the  segment term for a path, resulting in unbiased results and faster convergence compared to existing models. \revise{Furthermore, we also propose a novel PDF for BRDF multiple importance sampling, allowing for a better match with the multiple-bounce BRDFs, producing less noise than commonly used approximations.} 

To our best knowledge, our model is the first to use the invariance principle with an anisotropic phase function. By introducing the invariance principle to the multiple-bounce microfacet, we have acquired a simple derivation and a simple form. And we believe the invariance principle can be applied to similar problems, like the layered microflake (SpongeCake) model~\cite{Wang:2022:sponge}. Furthermore, the multiple bounces of a refractive surface is challenging, but is also important as future work for a complete understanding of the microfacet theory.



\begin{acks}
    We thank the reviewers for the valuable comments. This work has been partially supported by the National Key R\&D Program of China under grant No. 2022ZD0116305 and National Natural Science Foundation of China under grant No. 62172220.
\end{acks}



\bibliographystyle{ACM-Reference-Format}
\bibliography{paper}
\mycfigure{curve}{inverses.pdf}{Comparison between our model (PT), Heitz et al.~\shortcite{heitz2016}, \revise{Wang et al.~\shortcite{wang2021positionfree} (PT)}, and Bitterli and d'Eon~\shortcite{BitterliAndd'Eon:2022} (PT) in terms of BRDF reflectance and inverse efficiency (variance $\times$ the time cost) for the rough conductor with roughness 0.5 and 1.0. Note that we use the unidirectional estimator for all models. Here, we visualize both terms as a function of the outgoing direction, given an incoming direction. $\theta_i$ and $\theta_o$ are the angles between the incident/exit directions and the normal to the macrosurface, respectively. Our BRDF value can match both Heitz et al.~\shortcite{heitz2016} and Biltterli and d'Eon~\shortcite{BitterliAndd'Eon:2022}, while  our method is the most efficient in all cases (varying roughness or different incoming directions). Wang et al.~\shortcite{wang2021positionfree} can not match the groundtruth and introduces bias due to the independent-bounce assumption, although their model has higher performance than ours, when $\theta_o$ = 0 for $\alpha = 1.0$.
}

\mycfigure{sphere}{matpreview_difference_v3.pdf}{ Comparison between our models (PT and BDPT), Wang et al.~\shortcite{wang2021positionfree} (PT and BDPT) and Bitterli and d'Eon~\shortcite{BitterliAndd'Eon:2022} (PT) with equal sampling rate (4 spp) on the {Matpreview} scene. Our model (BDPT) produces the highest quality. Note that the model by Bitterli and d'Eon~\shortcite{BitterliAndd'Eon:2022} has the same variance as ours (PT) with an equal sampling rate, but has a higher time cost. The result of Wang et al.~\shortcite{wang2021positionfree} (BDPT) has a low noise level, but shows a larger error than Heitz et al.~\shortcite{heitz2016}, for the reason of bias. \revise{The difference images are shown on the bottom right of each image.}}

\mycfigure{pdf_pot}{pdf_pot_v5.pdf}{Comparison between our \revises{MIS} PDF and single bounce + the Lambertian term (Heitz et al.~\shortcite{heitz2016}) with equal sampling rate (16 spp) on the Pot scene \revises{(aluminium, $\alpha$ = 0.15, $\alpha$ = 0.2 and $\alpha$ = 0.1)}. \revises{We use the same BRDF evaluation and sample in both results, and the only difference is the PDF used for MIS.} }

\mycfigure{pdfCurve}{pdfcurve_line.pdf}{\revise{Comparison between our improved \revises{MIS} PDF, the previous solution (single bounce + the Lambertian term) by Heitz et al.~\shortcite{heitz2016} and the ground truth using the multiple-bounce BRDF reflectance. \revises{The ground truth is plotted with a stochastic aggregate of multiple-bounce BRDF with 100K samples.} Our PDF matches the ground-truth PDF at all roughness, while the other solution~\cite{heitz2016} only fits well at high roughness.}}

\mycfigure{vase}{vase_v9.pdf}{\revise{Equal-time comparison (about 2.8 seconds) between our model (BDPT), Wang et al.~\shortcite{wang2021positionfree} (BDPT) and Heitz et al.~\shortcite{heitz2016} on the {Vase} scene. Our model produces the highest quality with equal time.}}

\mycfigure{linegraph}{line_graph_v3.pdf}{\revise{Left: The time cost of segment term evaluation as a function of the bounce count. Note that the time cost is almost linear to bounce.} \revises{Middle:} The error (MSE) of our methods (both BDPT and PT), Wang et al.~\shortcite{wang2021positionfree} and Heitz et al.~\shortcite{heitz2016} over varying sampling rate on the {Matpreview} scene (Figure~\ref{fig:sphere}). The converged result by Heitz et al.~\shortcite{heitz2016} is treated as the ground truth. Our models can converge to the ground truth, while the model by Wang et al.~\shortcite{wang2021positionfree} is biased. Both our models converge faster than Heitz et al.~\shortcite{heitz2016}, while the convergence rate of our model (BDPT) is the fastest. \revises{Right: The percentage of light paths as a function of the ray count pointing downwards on the Vase and DecorativeSet scenes.
}  }

\end{document}



\title{Supplemental materials: Multiple-bounce Smith Microfacet BRDFs using the Invariance Principle}


\author{Yuang Cui}
\orcid{0009-0006-8983-7844}
\authornote{Contribute equally. Research done when Yuang Cui was an intern at Nanjing University of Science and
Technology. }
\affiliation{%
  \institution{Anhui Science and Technology University}
  \country{China}
}
\email{yuangcui@outlook.com}

\author{Gaole Pan}
\orcid{0009-0007-9335-333X}
\authornotemark[1]
\affiliation{%
  \institution{Nanjing University of Science and Technology}
  \country{China}
}
\email{pangaole@njust.edu.cn}

\author{Jian Yang}
\orcid{0000-0003-4800-832X}
\affiliation{%
  \institution{Nanjing University of Science and Technology}
  \country{China}
}
\email{csjyang@njust.edu.cn}

\author{Lei Zhang}
\orcid{0000-0002-2078-4215}
\affiliation{%
  \institution{The Hong Kong Polytechnic University}
  \country{China}
}
\email{cslzhang@comp.polyu.edu.hk}

\author{Ling-Qi Yan}
\orcid{0000-0002-9379-094X}
\affiliation{%
  \institution{University of California Santa Barbara}
  \country{USA}
}
\email{lingqi@cs.ucsb.edu}

\author{Beibei Wang}
\orcid{0000-0001-8943-8364}
\authornote{Corresponding author.}
\affiliation{
    \institution{Nankai University, Nanjing University of Science and Technology}
    \country{China}
}
\email{beibei.wang@njust.edu.cn}

\renewcommand{\shortauthors}{Cui et al.}

\maketitle

\allowdisplaybreaks
\section{Relationship to the model by Bitterli and d'Eon~\cite{BitterliAndd'Eon:2022}}
\label{sec:Equivalence Analysis}

Our model and the model by Bitterli and d'Eon~\shortcite{BitterliAndd'Eon:2022} are related. They have different formulations, but their formulations can be proved equivalent for specific bounces.  In this section, we provide the formulation of these two models in Sec.~\ref{sec:formulation} and then prove their equivalence for specific bounces (2 and 3) in Sec. \ref{sec:equivalence}.

\subsection{Formulation of two models}
\label{sec:formulation}

\paragraph{Our model}
We follow the multiple-bounce BRDF model by Wang et al.  \shortcite{wang2021positionfree}, defining the contribution of a light path with vertex terms and segment terms. In our model, the contribution of a light path includes vertex terms and a segment term:
\begin{equation}
    f(\bar x) = \left(\prod_{i=0}^{i=k-1} v_i\right) S_k(\bd_0,\bd_1,\dots,\bd_k),
    \label{eq:lightpath}
\end{equation}
where the segment term $S_k$ has the formulation, as also shown in the main text:
\begin{equation}
    \begin{aligned}
S_1\left(\mathbf{d}_0,  \mathbf{d}_1\right) & =\frac{1}{1+\Lambda (-\mathbf{d}_0)+\Lambda (\mathbf{d}_1)},  \\
S_k\left(\mathbf{d}_0, \dots , \mathbf{d}_k\right) & = \begin{cases}
0,\text { if } \mathbf{d}_{0} . z>0\text { or }\mathbf{d}_{k} . z<0,\\
S_1\left(\mathbf{d}_0,  \mathbf{d}_k\right)\left ( S{_{k-1}}\left(\mathbf{d}_0, \dots ,  \mathbf{d}_{k-1}\right)+\right.\\
\left.S_{k-1}\left(\mathbf{d}_1, \dots ,  \mathbf{d}_k\right) \right ),\text{ otherwise}. \\
\end{cases}\\
\end{aligned}
\end{equation}

\paragraph{The model by Bitterli and d'Eon~\shortcite{BitterliAndd'Eon:2022}}
Bitterli and d'Eon \shortcite{BitterliAndd'Eon:2022} perform a preintegration of the height component, using an explicit formulation of the height distribution as a hyperexponential distribution, resulting in an angular domain-only formulation. Their final formulation includes a phase function and a $p_\mathrm{exit}$:
\begin{equation}
    f(\bar x) = \left(\prod_{i=0}^{i=k-1} f_p(\bd_i,\bd_{i+1})\right) p_\mathrm{exit}(\bd_0,\bd_1,\dots,\bd_k),
\end{equation}
where $f_p$ is the phase function of the scattering, as defined by Heitz et al.~\shortcite{heitz2016}: 
\begin{equation}
    f_p =  \frac{F \left(\omega_i, \omega_h\right) D_{\omega_i} \left(\omega_h \right)}{4 \left\vert \omega_h \cdot \omega_i \right \vert}.
\end{equation}

The $p_\mathrm{exit}$ is the probability of a photon exiting the medium, conditioned on the directions it takes after each collision, defined as:
\begin{equation}
p_\mathrm{exit}(\bd_0,\bd_1,\dots,\bd_k)=\sum_{j=1}^{N_{i}} \frac{a_{i, j}}{b_{i, j}+\Lambda(\bd_k)}.
\end{equation}

\begin{algorithm}[h]
\caption{SegmentTerm}
\label{alg:SegmentTerm}
\Class(\tcp*[h]{Initialize $\Lambda_o$}){$\mathrm{SegmentTerm}(\Lambda_o)$} {

    \Def{$m \gets 1$} \;
    \Def{$N \gets 0$} \;
    \Def{$e[], g[], l[]$} \;

    \Fn{$\mathrm{addBounce}(\Lambda_k)$}{
       \eIf(\tcp*[h]{Downwards}){$\Lambda_k < 0$} {
        $N \gets N+1$\;
        $l[N] \gets |\Lambda_k|$\;
        $e[N] \gets \frac{1}{\Lambda_o+|\Lambda_k|}$\;
        $g[N] \gets 0$\;
        $m \gets m \cdot e[N]$\Comment{cache the results when sampled rays are only downwards}
        }(\tcp*[h]{Upwards}){
        
            \eIf{$m = 0$}{
               $g[N] \gets \frac{g[N]}{\Lambda_k + l[N]}$\;
        }(\tcp*[h]{Initialize g}){
               $g[N] \gets \frac{1}{\Lambda_k + l[N]}$\;
               $m \gets 0$\;
            }

            \For{$i \gets N-1 ... 1$}{
                $g[i] \gets \frac{g[i] + g[i + 1]}{\Lambda_k + l[i]}$\;
            }
        }
    }

    \Fn{$\mathrm{getS_k}()$}{

        \If(\tcp*[h]{Return results if cached}){$m \ne 0$}{
        \KwRet{ m }\;
        }
         $s \gets 0$\;
        \For{$i \gets N ... 1$}{
            $s \gets e[i] \cdot (s + g[i])$\;
        }
        \KwRet{ s }\;
    
    }
}

\end{algorithm}

$N_{i}, a_{i, j}, b_{i, j}$ are the number and coefficients of exponentials in the i-th height distribution, where:
\begin{equation}
\begin{aligned}
 a_{1,1}&=b_{1,1}=\Lambda(\bd_0),N_{1}=1,\\
a_{i+1, j}^{\downarrow} & =\left\{\begin{array}{ll}
a_{i, j} \frac{\Lambda(\bd_{i})}{\Lambda(\bd_{i})-b_{i, j}}, & \text { if } j<N_{i+1}, \\
\sum_{j=1}^{N_{i}}-a_{i+1, j}, & \text { else, }
\end{array}\right. \\
b_{i+1, j}^{\downarrow} & =\left\{\begin{array}{ll}
b_{i, j}, & \text { if } j<N_{i+1}, \\
\Lambda(\bd_{i}), & \text { else, }
\end{array} \text { and } N_{i+1}^{\downarrow}=N_{i}+1,\right.\\
a_{i+1, j}^{\uparrow}&=a_{i, j} \frac{\Lambda(\bd_{i})}{b_{i, j}+\Lambda(\bd_{i})}, b_{i+1, j}^{\uparrow}=b_{i, j} \text { and } N_{i+1}^{\uparrow}=N_{i}.
\end{aligned}
\label{eq:bitterli}
\end{equation}
The superscripted arrow is the direction of the i-th bounce.

Note that, their model model might have singularities, which come from the minus of two $\Lambda$ functions.

\paragraph{Comparison between two models}
\sloppy
This $p_\mathrm{exit}$ from Bitterli and d'Eon~\shortcite{BitterliAndd'Eon:2022} is similar but not equivalent to our segment term, since some terms in their $p_\mathrm{exit}$ are included in our vertex term. The relationship between our path segment term and the $p_\mathrm{exit}$ is:
\begin{equation}
    p_\mathrm{exit}(\bd_0,\bd_1,\dots,\bd_k)= \left(\prod_{i=0}^{i=k-1}\left|\Lambda(\bd_i)\right|\right)S_k(\bd_0,\bd_1,\dots,\bd_k).
\end{equation}

\subsection{Equivalence derivation between two models}
\label{sec:equivalence}
We find that Bitterli and d'Eon~\shortcite{BitterliAndd'Eon:2022}'s model and ours are equivalent for specific bounces. Here, we provide explicit derivations of equivalence between their model and ours for specific paths with 2 or 3 bounces. However, it is non-trivial to generalize to arbitrary bounces.

\paragraph{Equivalence derivation for light path with two bounces.}

\begin{align*}
\begin{aligned}
&p_\mathrm{exit}(\bd_0,\bd_1,\bd_2)=\sum_{j=1}^{N_{i}} \frac{a_{i, j}}{b_{i, j}+\Lambda(\bd_k)}\\
=&\frac{|\Lambda(\mathbf{d}_0)|\cdot  \frac{|\Lambda(\mathbf{d}_1)|}{|\Lambda(\mathbf{d}_1)|-|\Lambda(\mathbf{d}_0)|} }{\Lambda(\mathbf{d}_2 )+|\Lambda(\mathbf{d}_0)|} + \frac{-|\Lambda(\mathbf{d}_0)|  \cdot  \frac{|\Lambda(\mathbf{d}_1)|}{|\Lambda(\mathbf{d}_1)|-|\Lambda(\mathbf{d}_0)|}}{\Lambda(\mathbf{d}_2 )+|\Lambda(\mathbf{d}_1)|} 
\\=&  \frac{|\Lambda(\mathbf{d}_0)|\cdot|\Lambda(\mathbf{d}_1)|}{|\Lambda(\mathbf{d}_1)|-|\Lambda(\mathbf{d}_0)|}\left ( \frac{1}{\Lambda(\mathbf{d}_2 )+|\Lambda(\mathbf{d}_0)|}-\frac{1}{\Lambda(\mathbf{d}_2 )+|\Lambda(\mathbf{d}_1)|} \right ) 
\\=&  \frac{|\Lambda(\mathbf{d}_0)|\cdot|\Lambda(\mathbf{d}_1)|}{|\Lambda(\mathbf{d}_1)|-|\Lambda(\mathbf{d}_0)|}\left ( \frac{\Lambda(\mathbf{d}_2 )+|\Lambda(\mathbf{d}_1)|-(\Lambda(\mathbf{d}_2 )+|\Lambda(\mathbf{d}_0)|)}{(\Lambda(\mathbf{d}_2 )+|\Lambda(\mathbf{d}_0)|)(\Lambda(\mathbf{d}_2 )+|\Lambda(\mathbf{d}_1)|)} \right ) 
\\=&  \frac{|\Lambda(\mathbf{d}_0)|\cdot|\Lambda(\mathbf{d}_1)|}{|\Lambda(\mathbf{d}_1)|-|\Lambda(\mathbf{d}_0)|}\left ( \frac{|\Lambda(\mathbf{d}_1)|-|\Lambda(\mathbf{d}_0)|}{(\Lambda(\mathbf{d}_2 )+|\Lambda(\mathbf{d}_0)|)(\Lambda(\mathbf{d}_2 )+|\Lambda(\mathbf{d}_1)|)} \right ) 
\\=&  \frac{|\Lambda(\mathbf{d}_0)|\cdot|\Lambda(\mathbf{d}_1)|}{(\Lambda(\mathbf{d}_2 )+|\Lambda(\mathbf{d}_0)|)(\Lambda(\mathbf{d}_2 )+|\Lambda(\mathbf{d}_1)|)}
\\=&\left(\prod_{i=0}^{i=k-1}\left|\Lambda(\bd_i)\right|\right)S_k(\bd_0,\bd_1,\bd_2).
\end{aligned}
\end{align*}

\begin{algorithm}[h]
\caption{Sample}
\label{alg:sample}
\KwResult{weight $w$ and direction $\bd_k$}
\BlankLine
$p \gets 1$  \Comment{pdf}
$k \gets 0$  \Comment{bounce}
$w \gets 1$  \Comment{weight}
$\bd_0 \gets -\omega_i$\;
\While{$\mathrm{true}$}{
   ($\bd_k$,$p_k$,$w_k$) $ \gets $ sample($ \bd_{k-1} $)\;
   $w \gets w \cdot w_k$\;
   $p \gets p \cdot p_k$\;
    \If{$\bd_k.z > 0$}{
    $G_1 \gets \frac{1}{|\Lambda(\bd_k)|}$\;
        \eIf(\tcp*[h]{the ray continues the tracing with $G_1$ as the probablity}){$\mathrm{rand()}<G_1$}
        {
            $p \gets p \cdot G_1$\;
            break;
        }{
            $p \gets p \cdot (1-G_1)$\;
        }
    }
$k \gets k+1$\;
    \If{$k \ge maxBounce$}{
        break;
    }
}
   $w \gets w \cdot S_k(\bd_0,...,\bd_k)$\;
   $w \gets {w \over p}$\;
\end{algorithm}

\begin{algorithm}[h]
\caption{Eval}
\label{alg:eval}
\Fn{$\mathrm{eval}(\omega_i,\omega_o)$}{
$\mathrm{SegmentTerm} \  s(\Lambda(\omega_o))$\Comment{ initializing }
$f \gets 0$\Comment{result}
$k \gets 0$ \Comment{bounce}
$w \gets 1$\Comment{weight}
$\bd_0 \gets -\omega_i$\;
\While{$\mathrm{true}$}{
    $s.\mathrm{addBounce}(\Lambda(\bd_{k}))$\;
    $k \gets k + 1$\;
    \If{$k >= rrDepth$}{
        \If{$\mathrm{rand}()>q$}{
        break;
        }
    $w \gets {w \over q}$\Comment{russian roulette}
    }
    $f \gets f + w \cdot v(\bd_{k},\omega_o)\cdot s.\mathrm{getS_k}() $\;
    \If{$k \ge maxBounce$}{
        break;
    }
    ($\bd_{k+1}$,$p_k$) $ \gets $ sample($ \bd_{k} $)\;
    $w \gets {w \cdot v(\bd_{k},\bd_{k+1}) \over p_k}$\;
}
        \KwRet{ f }\;
}
\end{algorithm}

\paragraph{Equivalence derivation for light path with three bounces.}
\begin{align*}
&p_\mathrm{exit}(\bd_0,\bd_1,\bd_2,\bd_3)=\sum_{j=1}^{N_{i}} \frac{a_{i, j}}{b_{i, j}+\Lambda(\bd_k)}\\
&\frac{|\Lambda(\mathbf{d}_0)|\cdot  \frac{|\Lambda(\mathbf{d}_1)|}{|\Lambda(\mathbf{d}_1)|-|\Lambda(\mathbf{d}_0)|}\cdot  \frac{\Lambda(\mathbf{d}_2)}{\Lambda(\mathbf{d}_2)+|\Lambda(\mathbf{d}_0)|} }{\Lambda(\mathbf{d}_3 )+|\Lambda(\mathbf{d}_0)|} 
\\+& \frac{-|\Lambda(\mathbf{d}_0)|  \cdot  \frac{|\Lambda(\mathbf{d}_1)|}{|\Lambda(\mathbf{d}_1)|-|\Lambda(\mathbf{d}_0)|}\cdot  \frac{\Lambda(\mathbf{d}_2)}{\Lambda(\mathbf{d}_2)+|\Lambda(\mathbf{d}_1)|}}{\Lambda(\mathbf{d}_3 )+|\Lambda(\mathbf{d}_1)|}
\\=&\frac{{|\Lambda(\mathbf{d}_0)|\cdot|\Lambda(\mathbf{d}_1)|\cdot|\Lambda(\mathbf{d}_2)|}}{\left(|\Lambda(\mathbf{d}_1)|-|\Lambda(\mathbf{d}_0)|\right)\left(|\Lambda\left(\mathbf{d}_{1}\right)|+\Lambda\left(\mathbf{d}_{2}\right)\right)}
\\\cdot&\frac{1}{\left(|\Lambda\left(\mathbf{d}_{1}\right)|+\Lambda\left(\mathbf{d}_{3}\right)\right)\left(|\Lambda\left(\mathbf{d}_{0}\right)|+\Lambda\left(\mathbf{d}_{2}\right)\right)\left(|\Lambda\left(\mathbf{d}_{0}\right)|+\Lambda\left(\mathbf{d}_{3}\right)\right)}
\\\cdot&\left(
|\Lambda\left(\mathbf{d}_{0}\right)|\cdot\Lambda\left(\mathbf{d}_{3}\right)-
|\Lambda\left(\mathbf{d}_{0}\right)|\cdot\Lambda\left(\mathbf{d}_{2}\right)+
|\Lambda\left(\mathbf{d}_{0}\right)|^2\right.
\\-&\left.|\Lambda\left(\mathbf{d}_{1}\right)|\cdot\Lambda\left(\mathbf{d}_{3}\right)+
|\Lambda\left(\mathbf{d}_{1}\right)|\cdot\Lambda\left(\mathbf{d}_{2}\right)-
|\Lambda\left(\mathbf{d}_{1}\right)|^2\right)
\\=&{\tiny \ \ \ } \frac{|\Lambda(\mathbf{d}_0)|\cdot|\Lambda(\mathbf{d}_1)|\cdot\Lambda(\mathbf{d}_2)}{\left(|\Lambda\left(\mathbf{d}_{1}\right)|+\Lambda\left(\mathbf{d}_{2}\right)\right)\left(|\Lambda\left(\mathbf{d}_{0}\right)|+\Lambda\left(\mathbf{d}_{3}\right)\right)} \\
\cdot&\left ( \frac{1}{|\Lambda\left(\mathbf{d}_{0}\right)|+\Lambda\left(\mathbf{d}_{2}\right)}+\frac{1}{|\Lambda\left(\mathbf{d}_{1}\right)|+\Lambda\left(\mathbf{d}_{3}\right)} \right )
\\=&\left(\prod_{i=0}^{i=k-1}\left|\Lambda(\bd_i)\right|\right)S_k(\bd_0,\bd_1,\bd_2,\bd_3).
\end{align*}

\begin{algorithm}[h]
\caption{SegmentTerm (Ours)}
\label{alg:coloredST}
\Class{$\mathrm{SegmentTerm}(\Lambda_o)$} {

    $N \gets 0$ \;
    $m \gets 1$ \;
    \Def{$e[], g[], l[]$} \;

    \Fn{$\mathrm{addBounce}(\Lambda_k)$}{
       \eIf{$\Lambda_k < 0$} {
      \mybox{green}{24mm}{
       $N \gets N+1$\;
        $l[N] \gets |\Lambda_k|$\;
        $e[N] \gets \frac{1}{\Lambda_o+|\Lambda_k|}$\;
        $g[N] \gets 0$\;
        $m \gets m \cdot e[N]$\;
        }
        }{
        
    \mybox{yellow}{33.5mm}{
            \eIf{$m = 0$}
            {
               $g[N] \gets \frac{g[N]}{\Lambda_k + l[N]}$\;
        }{
               $g[N] \gets \frac{1}{\Lambda_k + l[N]}$\;
               $m \gets 0$\;
            }
\mybox{red}{32mm}{
            \For{$i \gets N-1 ... 1$}{
                $g[i] \gets \frac{g[i] + g[i + 1]}{\Lambda_k + l[i]}$\;
            }}
            }
        }
    }

    \Fn{$\mathrm{getS_k}()$}{
    \mybox{cyan}{33.5mm}{
        \If{$m \ne 0$}{
        \KwRet{ m }\;
        }
         $s \gets 0$\;
         
\mybox{red}{32mm}{
        \For{$i \gets N ... 1$}{
            $s \gets e[i] \cdot (s + g[i])$\;
        }}
        \KwRet{ s }\;
    }
    }
}

\end{algorithm}




\section{Implementation details}
\label{sec:impl}

In this section, we provide the pseudo-code for our models (both BRDF evaluation and sample) and show the explicit implementation difference between Bitterli and d'Eon~\shortcite{BitterliAndd'Eon:2022} and our model.

\subsection{Implementation details of our model}



\paragraph{Evaluation}
We provide the pseudo-code of our BRDF evaluation with the unidirectional estimator in Alg.~\ref{alg:eval} and further show the detailed implementation of our segment term using dynamic programming in Alg.~\ref{alg:SegmentTerm}.

\begin{algorithm}[h]
\caption{HeightDistribution (Bitterli and d'Eon)}
\label{alg:coloredHD}
\Class{$\mathrm{HeightDistribution(\Lambda_o)}$} {

    \Def{$N, a[], b[]$} \;

    \Fn{$\mathrm{addBounce}(\Lambda_k)$}{
    \uIf{$N=0$}{
    $a[1] \gets \Lambda_k$\;
    $b[1] \gets \Lambda_k$\;
    $N \gets N+1$\;
    }
    \uElseIf{$\Lambda_k < 0$} {
    \mybox{green}{41.5mm}{
    \mybox{red}{40mm}{
        \For{$i \gets 1 ... N$}{
            $a[i] \gets a[i]\frac{\Lambda_k}{\Lambda_k-b[i]} $\;
        }
        }
        $a[N+1] \gets 0$\;
        \mybox{red}{40mm}{
        \For{$i \gets 1 ... N$}{
            $a[N+1] \gets a[N+1]-a[i] $\;
        }
        }
        $b[N+1] \gets \Lambda_k$\;
        $N \gets N+1$\;
    }
    }
    \Else{
    \mybox{yellow}{33.5mm}{
    \mybox{red}{32mm}{
        \For{$i \gets 1 ... N$}{
            $a[i] \gets a[i]\frac{\Lambda_k}{\Lambda_k+b[i]} $\;
        }
        }
        $a[N+1] \gets 0$\;
        $b[N+1] \gets -\Lambda_k$\;}
    }
}
    \Fn{$\mathrm{getP_{exit}}()$}{
    \mybox{cyan}{28.5mm}{
         $s \gets 0$\;
    \mybox{red}{27mm}{
        \For{$i \gets 1 ... N$}{
            $s \gets s + \frac{a[i]}{\Lambda_o+b[i]}$\;
        }
        }
        \KwRet{ s }\;
        }
    
    }
}

\end{algorithm}

\paragraph{Sample}

We show the multiple-bounce BRDF sample in Alg. \ref{alg:sample}. At each bounce, we sample the VNDF to get the outgoing direction and then compute the masking function ($G_1$ function) of the sampled ray to decide whether to exit the microgeometry. Otherwise, we treat the sampled direction as the incoming direction and continue sampling until the ray leaves the surface. After getting such a light path, we compute the path contribution by evaluating the light path divided by the PDF. Using the masking function of the last sampled ray as the exit probability enables satisfying results with less time cost, while using the entire sampled path to compute the exit probability introduces a large time overhead.

\paragraph{BDPT}

The bidirectional estimator is the same as Wang et al.~\shortcite{wang2021positionfree}, except for the segment term. We did not provide the results of Bitterli and d'Eon \shortcite{BitterliAndd'Eon:2022} (BDPT), since we can not get the correct results after implementing their pseudo-code. We also find that their algorithm does not provide an expected result (single-bounce microfacet BRDF, even set the path length as 2).

\subsection{Implementation comparison between our model and Bitterli and d'Eon}

In this section, we compare the implementation details between our segment term and the $p_\mathrm{exit}$ by Bitterli and d'Eon~\shortcite{BitterliAndd'Eon:2022}. We have highlighted the main parts of our algorithm in Alg. \ref{alg:coloredST} and their algorithm in Alg. \ref{alg:coloredHD}. 

The different formulations of the two algorithms lead to different performances. In the end, our method has less time cost when rendered with an equal sample rate.


\section{More results}
As shown in Figure~\ref{fig:anisocurve}, our PDF is less accurate for anisotropic BRDFs than for isotropic media, since our mapping function averages the roughness for anisotropic BRDFs. Even though, our PDF is still better than the single bounce + the Lambertian term approximation.
\myfigure{anisocurve}{pdfcurve_aniso.pdf}{Comparison between our improved PDF, the previous solution (single bounce + the Lambertian term) by Heitz et al.~\shortcite{heitz2016} and the ground truth using the multiple-bounce BRDF reflectance.}
\bibliographystyle{ACM-Reference-Format}
\bibliography{paper}